\documentclass[prx,superscriptaddress,amsmath,amssymb,twocolumn]{revtex4}
\usepackage{natbib}
\usepackage{graphicx}         
\usepackage{dcolumn}          
\usepackage{bm}               
\usepackage{upgreek}
\usepackage{booktabs} 				
\usepackage{tabularx}
\usepackage{array}
\usepackage[colorlinks=true,urlcolor=blue,breaklinks=true]{hyperref}
\usepackage{xcolor}
\usepackage{float}

\begin{document}

\title{Hard-gap spectroscopy in a self-defined mesoscopic InAs/Al nanowire Josephson junction}

\author{Patrick Zellekens}
\email{p.zellekens@fz-juelich.de}
\affiliation{Peter Gr\"unberg Institut (PGI-9), Forschungszentrum J\"ulich, 52425 J\"ulich, Germany}
\affiliation{JARA-Fundamentals of Future Information Technology, J\"ulich-Aachen Research Alliance, Forschungszentrum J\"ulich and RWTH Aachen University, Germany}

\author{Russell Deacon}
\affiliation{RIKEN Center for Emergent Matter Science and Advanced Device Laboratory, 351-0198 Saitama, Japan}

\author{Pujitha Perla}
\affiliation{Peter Gr\"unberg Institut (PGI-9), Forschungszentrum J\"ulich, 52425 J\"ulich, Germany}
\affiliation{JARA-Fundamentals of Future Information Technology, J\"ulich-Aachen Research Alliance, Forschungszentrum J\"ulich and RWTH Aachen University, Germany}

\author{H. Aruni Fonseka}
\affiliation{Department of Physics, University of Warwick, Coventry
CV4 7AL, UK}

\author{Timm Moerstedt}
\affiliation{Peter Gr\"unberg Institut (PGI-9), Forschungszentrum J\"ulich, 52425 J\"ulich, Germany}
\affiliation{JARA-Fundamentals of Future Information Technology, J\"ulich-Aachen Research Alliance, Forschungszentrum J\"ulich and RWTH Aachen University, Germany}

\author{Steven A. Hindmarsh}
\affiliation{Department of Physics, University of Warwick, Coventry
CV4 7AL, UK}

\author{Benjamin Bennemann}
\affiliation{Peter Gr\"unberg Institut (PGI-9), Forschungszentrum J\"ulich, 52425 J\"ulich, Germany}
\affiliation{JARA-Fundamentals of Future Information Technology, J\"ulich-Aachen Research Alliance, Forschungszentrum J\"ulich and RWTH Aachen University, Germany}

\author{Florian Lentz}
\affiliation{Helmholtz Nano Facility, Forschungszentrum J\"ulich, 52425 J\"ulich, Germany}

\author{Mihail I. Lepsa}
\affiliation{Peter Gr\"unberg Institut (PGI-10), Forschungszentrum J\"ulich, 52425 J\"ulich, Germany}
\affiliation{JARA-Fundamentals of Future Information Technology, J\"ulich-Aachen Research Alliance, Forschungszentrum J\"ulich and RWTH Aachen University, Germany}

\author{Ana M. Sanchez}
\affiliation{Department of Physics, University of Warwick, Coventry
CV4 7AL, UK}

\author{Detlev Gr\"utzmacher}
\affiliation{Peter Gr\"unberg Institut (PGI-9), Forschungszentrum J\"ulich, 52425 J\"ulich, Germany}
\affiliation{Peter Gr\"unberg Institut (PGI-10), Forschungszentrum J\"ulich, 52425 J\"ulich, Germany}
\affiliation{JARA-Fundamentals of Future Information Technology, J\"ulich-Aachen Research Alliance, Forschungszentrum J\"ulich and RWTH Aachen University, Germany}

\author{Koji Ishibashi}
\affiliation{RIKEN Center for Emergent Matter Science and Advanced Device Laboratory, 351-0198 Saitama, Japan}

\author{Thomas Sch\"apers}
\affiliation{Peter Gr\"unberg Institut (PGI-9), Forschungszentrum J\"ulich, 52425 J\"ulich, Germany}
\affiliation{JARA-Fundamentals of Future Information Technology, J\"ulich-Aachen Research Alliance, Forschungszentrum J\"ulich and RWTH Aachen University, Germany}

\hyphenation{}
\date{\today}

\begin{abstract}
Superconductor/semiconductor-nanowire hybrid structures can serve as versatile building blocks to realize Majorana circuits or superconducting qubits based on quantized levels such as Andreev qubits. For all these applications it is essential that the superconductor-semiconductor interface is as clean as possible. Furthermore, the shape and dimensions of the superconducting electrodes needs to be precisely controlled. We fabricated self-defined InAs/Al core/shell nanowire junctions by a fully in-situ approach, which meet all these criteria. Transmission electron microscopy measurements confirm the sharp and clean interface between the nanowire and the in-situ deposited Al electrodes which were formed by means of shadow evaporation. Furthermore, we report on tunnel spectroscopy, gate and magnetic field-dependent transport measurements. The achievable short junction lengths,the observed hard-gap and the magnetic field robustness make this new hybrid structure very attractive for applications which rely on a precise control of the number of sub-gap states, like Andreev qubits or topological systems.
\end{abstract}
\maketitle

\section{Introduction}

Over the last decade, superconductor-semiconductor nanowire hybrid structures have experienced growing interest as building blocks in circuits based on various novel physical phenomena. In superconducting transmon quantum bits Josephson junctions featuring a semiconductor nanowire weak link between two superconducting electrodes allow for gate control instead of flux control as employed in conventional Josephson junctions \cite{Larsen15,deLange15,Casparis16,Luthi18}. This makes the circuit layout much more compact. In addition, one can also make use of the relatively large Fermi wavelength in the semiconductor being comparable with the nanowire dimensions. As a consequence, well-distinguished quantized levels are formed, which might be used to realize an Andreev quantum bit \cite{Zazunov03,Woerkom17,Tosi19,Hays18}.

One of the most exciting properties of such nanowire-superconductor hybrid structures is the access to the mesoscopic regime, in which the behavior of the realized Josephson junction is fully-determined by the coherent bound states inside the semiconductor. Due to the presence of spin-orbit coupling in III-V semiconductor nanowires, it is also possible to observe exotic phenomena beyond the classical Andreev spectrum \cite{Gharavi15,Park17,VanDam06,Woerkom17,Tosi19}. After the first observation of signatures of Majorana fermions in nanowire-superconductor hybrid structures \cite{Mourik12,Deng12,Rokhinson12,Zhang17}, the interest has extended towards realizing topological quantum bits \cite{Albrecht16,Karzig17}. One reason is that owing to the non-local nature of Majorana states, topological quantum circuits are less prone to errors \cite{Alicea11}.
\begin{figure}[!ht]
\centering
\includegraphics[width=0.40\textwidth]{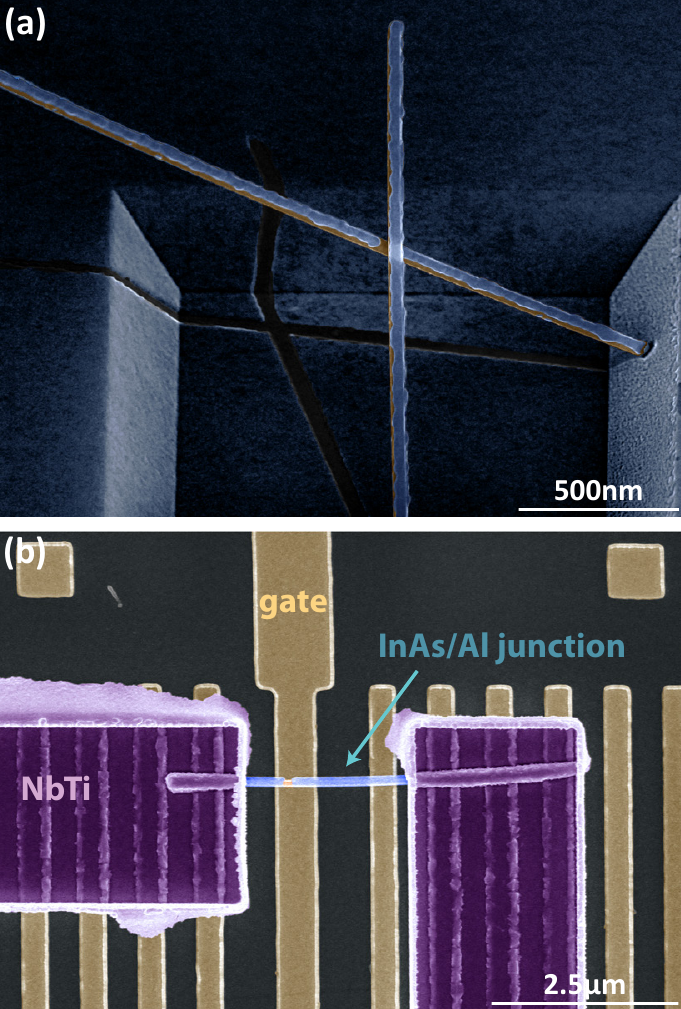}
\caption{(a) False-color scanning electron beam microscopy image of the as-grown InAs nanowires (orange) covered with an Al half-shell (blue). The in-situ junction is formed due to the shadow imposed by the upper nanowire. 
(b) A single Al/InAs nanowire junction contacted by NbTi electrodes. The junction was placed on a Ti/Au bottom gate electrode by means of a micromanipulator. A stack of Al$_2$O$_3$/HfO$_2$ was used as gate dielectric. }
\label{fig:circuit}
\end{figure}

The performance of nanowire-based Josephson junctions depends to a large extent on the properties of the superconductor-semiconductor interface. In most cases, the superconductor is deposited ex-situ after the growth of the nanowire. This makes it necessary to carry out some surface cleaning steps, e.g. wet chemical etching or Ar$^+$ sputtering, before superconductor deposition \cite{Guel17}. However, this procedure often results in a non-ideal interface. In order to circumvent this problem, efforts were undertaken to deposit the superconductor in-situ, i.e. without breaking the vacuum after the growth of the nanowires \cite{Krogstrup15,Chang15,Guesgen17,Gazibegovic17,Bjergfelt19}. Another crucial issue is the length of the weak link junction, since it determines the number of Andreev levels involved in the Josephson supercurrent. The conventional method to define a junction in an in-situ deposited superconducting shell wire has been to use electron beam lithography and wet chemical etching \cite{Larsen15}, which puts constraints on the minimum attainable junction length due to the isotropic nature of chemical etches. In order to tackle this issue, very recently, new fabrication schemes were developed where the junction length is geometrically defined by shadow evaporation \cite{Gazibegovic17,Schueffelgen19b,Carrad19}. 

We fabricated fully in-situ InAs nanowire-based Josephson junctions, which utilizes repetition shadow evaporation of Al to define the closely spaced superconducting electrodes. The structural properties of the InAs-Al core-halfshell nanowires were investigated using transmission electron microscopy. In the transport experiments, we were able to tune the junction from the fully depleted regime, i.e. Coulomb blockade, to a well-developed Josephson supercurrent range by biasing a lithographically defined bottom gate finger. Furthermore, we present voltage-driven measurements, which are sensitive to the density of states in the nanowire and the structure of the superconducting gap. Finally, we present an analysis of the junction behavior for externally applied magnetic fields parallel to the nanowire axis and out-of-plane. 

\section{Growth and device fabrication}
The fully in-situ Josephson junctions were fabricated by using the following procedure: In order to achieve adjacent Si(111) surfaces, anisotropic etching is employed with Tetramethyl ammonium hydroxide (TMAH) on a pre-patterned Si(100) substrate with arrays of 3\,$\mu$m squares. The resulted tilted planes have an angle of $54.7\,^\circ$ with respect to the unetched Si(100) surface. These Si(111) facets form the basis for the subsequent InAs nanowire growth by means of catalyst-free molecular beam epitaxy (MBE).

\begin{figure*}[!t]
    \centering
    \includegraphics[width=1.0\textwidth]{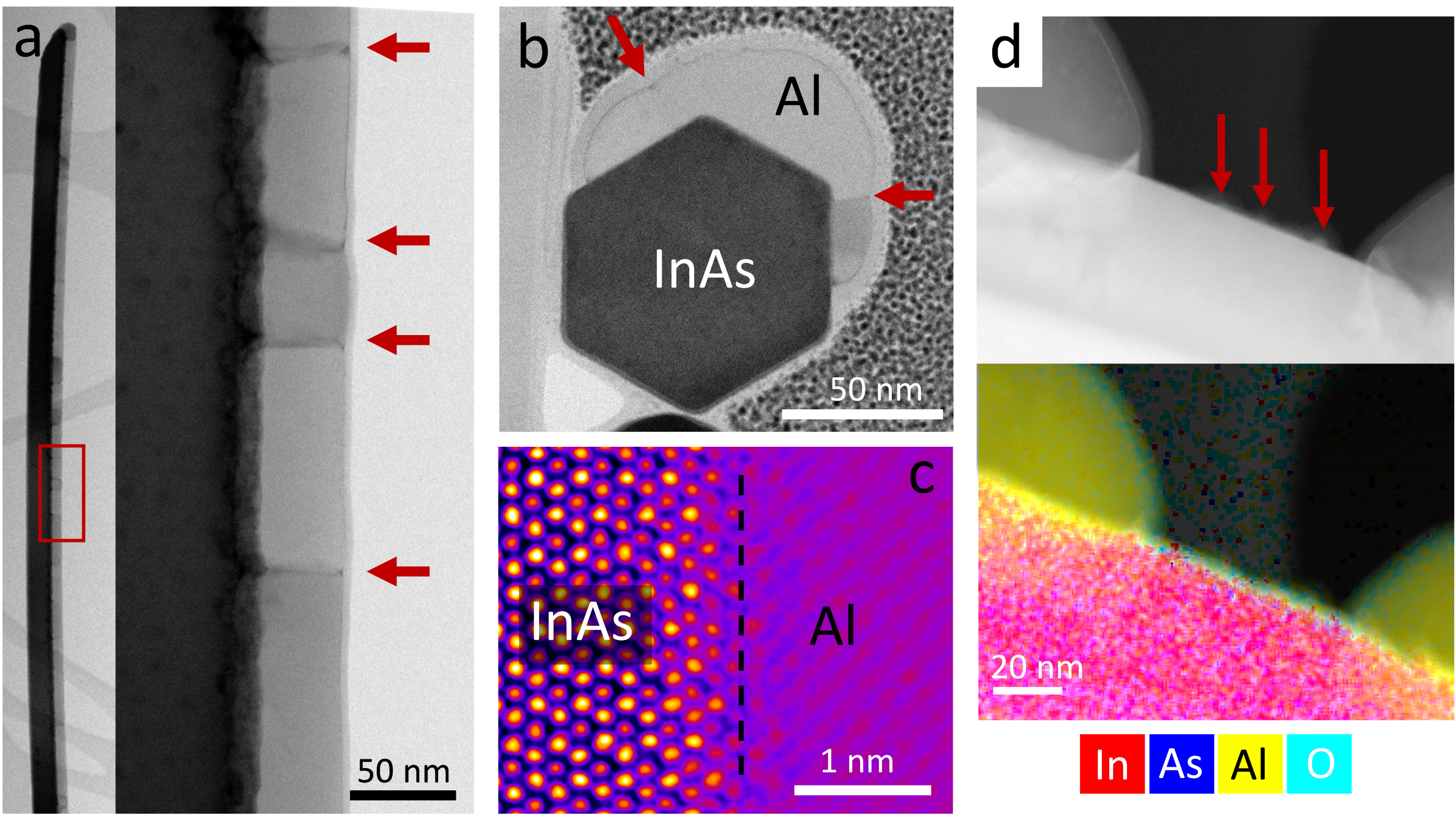}
    \caption{(a) A low magnification bright field (BF) scanning transmission electron microscope (STEM) image from the region indicated in the core-halfshell nanowire shown in the inset. Red arrows mark the grain boundaries. (b) A BF STEM image of a nanowires cross-section showing the Al half-shell and red arrows indicating the grain boundaries. (c) A false color high magnification image of the InAs-Al interface. The broken line marks the interface. (d) An annular dark field (ADF) STEM image and the corresponding EDX map of an InAs-Al junction showing the elemental distribution. The red arrows in the ADF image indicate the isolated oxidised Al droplets within the gap. }
    \label{fig:TEM-Al-InAs}
\end{figure*}

Before the NW growth, a 23$\,$nm thick SiO$_2$ layer is formed on the Si substrate by thermal oxidation. Then, 80$\,$nm wide  holes are defined in the oxide layer over adjacent Si(111) facets using electron beam (E-beam) lithography, reactive ion etching (RIE: CHF$_3$+O$_2$) and wet chemical  etching (HF).  The holes are positioned so that  the grown NWs cross each other closely, but do not merge. The InAs NWs are grown in two steps: First, at a substrate temperature of 480$\,^{\circ}$C with an In growth rate of 0.08$\,\mu$m/h (determined as the growth rate of In(100) planar layers) and an As$_4$ beam equivalent pressure (BEP) of approx. 4$\cdot$10$^{-5}\,$mbar for 10 min to sustain an optimal growth window and then decreasing the substrate temperature to 460$\,^{\circ}$C with an In growth rate of 0.03$\,\mu$m/h  and an As$_4$ BEP of approx. 3$\cdot$10$^{-5}\,$mbar for 3.5$\,$h, resulting in 4-5$\,\mu$m long and 80$\,$nm wide NWs. After the growth of the InAs nanowires, the substrate undergoes an arsenic desorption  at 400$\,^{\circ}$C for 20$\,$min and at 450$\,^\circ$C for 5$\,$min. By doing so, we suppress the formation of parasitic heterostructures like AlAs during the deposition of the Al, which ensures a pristine interface without any barriers. Subsequently, the sample is transferred to the metal MBE and is cooled down to $-6\,^\circ$C, followed by the evaporation of a 25$\,$nm thick layer of Al onto the nanowires at this temperature \cite{Krogstrup15,Guesgen17}. During the metal evaporation process, an  elevated angle of  87$^\circ$ is maintained between the metal source and the nanowire, which ensures that the metal deposition is smooth and highly crystalline. In Fig.~\ref{fig:circuit}(a) a scanning electron beam microscopy (SEM) image of an as-grown in-situ Al/InAs-nanowire Josephson junction with a junction length of approximately 80$\,$nm is shown. It can be clearly seen, that the junction length is directly related to the diameter of the upper nanowire, which acts as the shadow mask.

The devices were fabricated on commercially available highly resistive Si substrates ($\rho>100\,\mathrm{k}\Omega\mathrm{cm}$). All etching and metal-deposition steps were realized using standard e-beam lithography techniques. At first, a single Ti/Au (5$\,$nm/10$\,$nm) gate electrode surrounded by a set of electrically unconnected metal stripes for mechanical support were deposited, together with a small bonding pad and a single lead between the gate and the pad. This is followed by another Ti/Au deposition (60$\,$nm/70$\,$nm) to form a larger bonding pad on top of the smaller one as well as additional, positive e-beam markers. The whole substrate was subsequently covered by an 3$\,$nm/12$\,$nm thick Al$_2$O$_3$/HfO$_2$ dielectric layer by means of atomic layer deposition. As the circuits are intended to work for both AC and DC measurements, we use a transmission line to form the source contact of the device. The latter is terminated by an on-chip bias tee, consisting out of an interdigital capacitor and a planar coil. All three elements, together with the surrounding ground plane, were fabricated out of reactively sputtered titanium nitride (80 $\,$nm thick, deposited at room temperature).

Subsequently, the nanowires were then transfered onto the electrostatic gates by means of an SEM-based micro-manipulator setup. To ensure an ohmic coupling between the contacts, made out of NbTi, and the Al shell, we used a combination of a 5$\,$s long wet chemical etch in Transene-D, followed by an in-situ Ar$^{+}$ dry etch. The contact separation is chosen to be at least 1.5\,$\mu$m in order to reduce the effect of the wide-gap superconductor NbTi on the actual junction characteristics. Figure~\ref{fig:circuit}(b) shows one of the final Josephson junctions on top of a bottom gate electrode (yellow).

\section{Results}
\begin{figure*}[!t]
\centering
\includegraphics[width=0.99\textwidth]{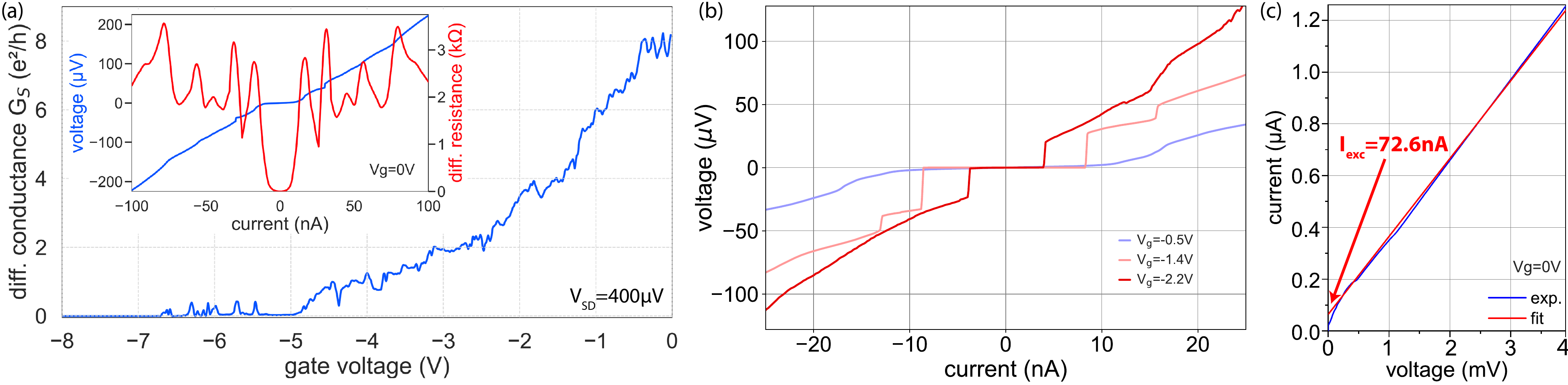}
\caption{Conductance $G$ vs. gate voltage measured in a voltage-driven setup with a constant voltage drop of 400\,$\mu$eV. Four different transport regimes can be distinguished (negative to positive gate voltage): Pinch-off, Coulomb blockade, tunnel-dominated transport, and classical Josephson junction behaviour. The inset shows the current voltage ($IV$) characteristics at zero gate voltage, with a supercurrent of approx. 21\,nA and signatures of subharmonic gap structures in the differential resistance. (b) IV curves at gate voltages of $-0.5$, $-1.4$, and $-2.2$\,V, respectively. (c) Current-voltage characteristics at $V_g=0$\,V up to large bias voltages. By fitting and extending the linear and ohmic behavior of the normal state an excess current $I_\mathrm{exc}$ of 72.6\,nA is extracted.}
\label{fig:gatesweep}
\end{figure*}
\subsection{Structural characterisation}
The structural properties of the InAs-Al core-half-shell nanowires were examined using scanning transmission electron microscopy (STEM). For this, the nanowires were transferred to holey carbon Cu grids by simply sweeping the arrays with the grid. The InAs nanowires grown by current catalyst-free method showed polytypic crystal structure containing thin, defective wurtzite (WZ) and zincblende (ZB) segments.

Figure~\ref{fig:TEM-Al-InAs} (a) shows a low magnification image of a core-halfshell nanowire. The Al layer is uniform along the nanowires and has a smooth surface with no significant faceting \cite{Krogstrup15}. Cross-sections prepared by focused ion beam revealed that the Al thickness is not uniform around the half-shell as shown in Fig.~\ref{fig:TEM-Al-InAs} (b). The maximum thicknesses of the Al layers in these nanowires are between 30--50\,nm. The Al half-shell is formed of large grains ($> 25$\,nm) with different orientations. These differently oriented grains can be seen in the two viewing directions as marked by red arrows in Figs.~\ref{fig:TEM-Al-InAs} (a) and (b). Some of the observed grain orientations include $([110]_{\textrm{Al}}\,||\,[0001]_{\textrm{InAs}})_{||} \times ([1-12]_{\textrm{Al}}\,||\,[11-20]_{\textrm{InAs}})_{\perp}$ and $([110]_{\textrm{Al}}\,||\, [0001]_{\textrm{InAs}})_{||} \times([1-10]_{\textrm{Al}} \, || \,[11-20]_{\textrm{InAs}})_{\perp}$ (using WZ notation for InAs). These orientations are different to those shown in some of the previous works \cite{Krogstrup15,Kang17,Carrad19}, but consistent with some of those seen in our prior results \cite{Guesgen17}. It should also be noted that the size of the Al grains is much larger than the width of a single crystal type region in the polytypic nanowire (which is generally less than 10 mono-layers), and these grains span multiple phase changes and defects. This means that change in phase of the nanowire does not directly induce change in metal grain orientation.

The interface between the Al and InAs was analysed using the cross sections, as this viewing angle avoids simultaneous contribution to the projection from Al growth on multiple facets. Fig.~\ref{fig:TEM-Al-InAs} (c) shows a falsely colorized higher magnification image of the InAs-Al interface. It can be seen that the interface is sharp and smooth with no amorphous material in between crystalline InAs and Al, as a result of fully in-situ deposition of the electrodes and absence of processing steps such as Ar$^+$  sputtering.  Edge dislocations which have been previously attributed to AlAs formation \cite{Krogstrup15, Guesgen17}, were observed in the side view of these nanowires. Dislocations were placed $\sim 2.5$\,nm from the interface within InAs. Evidence supporting an AlAs layer as thick as $2.5\,$nm was not seen in energy dispersive x-ray (EDX) data or lattice spacings of high resolution images, although a smaller lattice spacing ($d_{\{11-20\} }$) was observed between the edge-most lattice planes in some nanowire facets, accounting to a possible AlAs or AlInAs formation of thickness less than $0.5$\,nm. This shows that almost all of the As on the entire Nw surface has been desorbed prior to Al deposition and it is consistent with transport measurements (shown later) which show no evidence for the presence of a significant intermediate AlAs layer.
Figure~\ref{fig:TEM-Al-InAs} (d) shows an annular dark field (ADF) image and the corresponding EDX map of an InAs-Al junction. A clear sharp gap in Al layer is formed in the shadowed region with a junction width of 75\,nm. Few small, oxidised and isolated droplets of Al, which appears to have formed during its deposition are seen within the junction (indicated by red arrows in the ADF image). However, these seem to get etched-off during the subsequent device processing steps as no evidence of parallel  metallic bypass is observed in the transport measurements.

\subsection{Basic junction characteristics}
All presented measurements were performed in a He3/He4 dilution refrigerator at a base temperature of 15\,mK equipped with a superconducting magnet coil. $I-V$ traces are measured using a current bias supplied from a battery-powered current source and measured using a battery-powered differential voltage amplifier. The measurement lines within the refrigerator are heavily filtered with thermocoax, low-temperature copper powder filters and custom low pass filters in addition to room temperature 'pi'-filters. 

The efficiency of the gate response is one of the key properties of a nanowire junction. Therefore, we measured the conductance of the junction in a voltage-biased configuration. In  Fig.~\ref{fig:gatesweep} (a) a gate sweep for one of the in-situ devices is shown. For these measurements a constant voltage drop of 400\,$\mu$eV is maintained across the junction, i.e. above the induced proximity gap. Depending on the applied gate voltage $V_g$, four different regimes can be distinguished. For gate voltages below $-6.7$\,V, the nanowire is completely pinched off. Above this value, the current is mediated by single electron transport through an intrinsic quantum dot. For gate voltages larger than $-5\,$V, the nanowire opens up, but the conductance is still limited by a low-transparency tunnel coupling to the electron reservoirs in the contacts. This results in a conductance increase of just $2e^2/h$ over a gate voltage range of 2\,V and just a weakly pronounced superconducting branch. For the last section, i.e. voltages above $-3$\,V, the current voltage ($IV$) characteristics shows a classical Josephson junction response (cf. Fig. \ref{fig:gatesweep} (a) (inset)) with a clear supercurrent. In addition, signatures of subharmonic gap structures are found in the differential resistance \cite{Octavio83,Flensberg88}. 

Figure \ref{fig:gatesweep} (b) shows a set of IV curves at three different gate voltages within the Josephson-junction-regime. It can be seen, that the switching current of the device can be tuned between 3$\,$nA and 10$\,$nA. The additional constant-voltage-sections are related to self-induced Shapiro steps, caused by the transmission-line circuit on the sample. At $V_g=0$ we obtained a critical current of $I_c=21$\,nA and a normal state resistance of $R_N=3.33\,\mathrm{k}\Omega$, which results in an $I_cR_N$ product of $7\,\mu $V. We would like to stress that the presented $I_cR_N$ value just holds for the given gate voltage. Both the critical current as well as the normal state resistance are affected by quantum fluctuations, which are related to universal conductance fluctuations in the nanowire \cite{Guenel12,Doh05}. As indicated in Fig.~\ref{fig:gate_combined}(c), by extrapolating the current voltage characteristics at the normal state in the range $V > 400\,\mu$V a finite excess current $I_\mathrm{exc}$ of 72.6\,nA is extracted. The excess current can be used to get an estimation of the dominating type of transport within the channel by comparing the $I_\mathrm{exc}R_N$ product with the superconducting gap $\Delta$. Following the framework of the corrected Octavio--Tinkham--Blonder--Klapwijk theory \cite{Octavio83,Flensberg88}, we obtain a ratio of $eI_\mathrm{exc}R_N/\Delta=1.31$  at zero gate voltage. This value can be converted to the barrier strength parameter $Z=0.38$ and a corresponding contact transparency $\mathcal{T}=0.88$ \cite{Flensberg88,Niebler09}. This transparency is comparable to values obtained for hydrogen cleaned superconductor/nanowire junctions, i.e.  Al/InSb \cite{Gazibegovic17} or NbTiN/InSb \cite{Zhang17}.  
\begin{figure}[!t]
\centering
\includegraphics[width=0.49\textwidth]{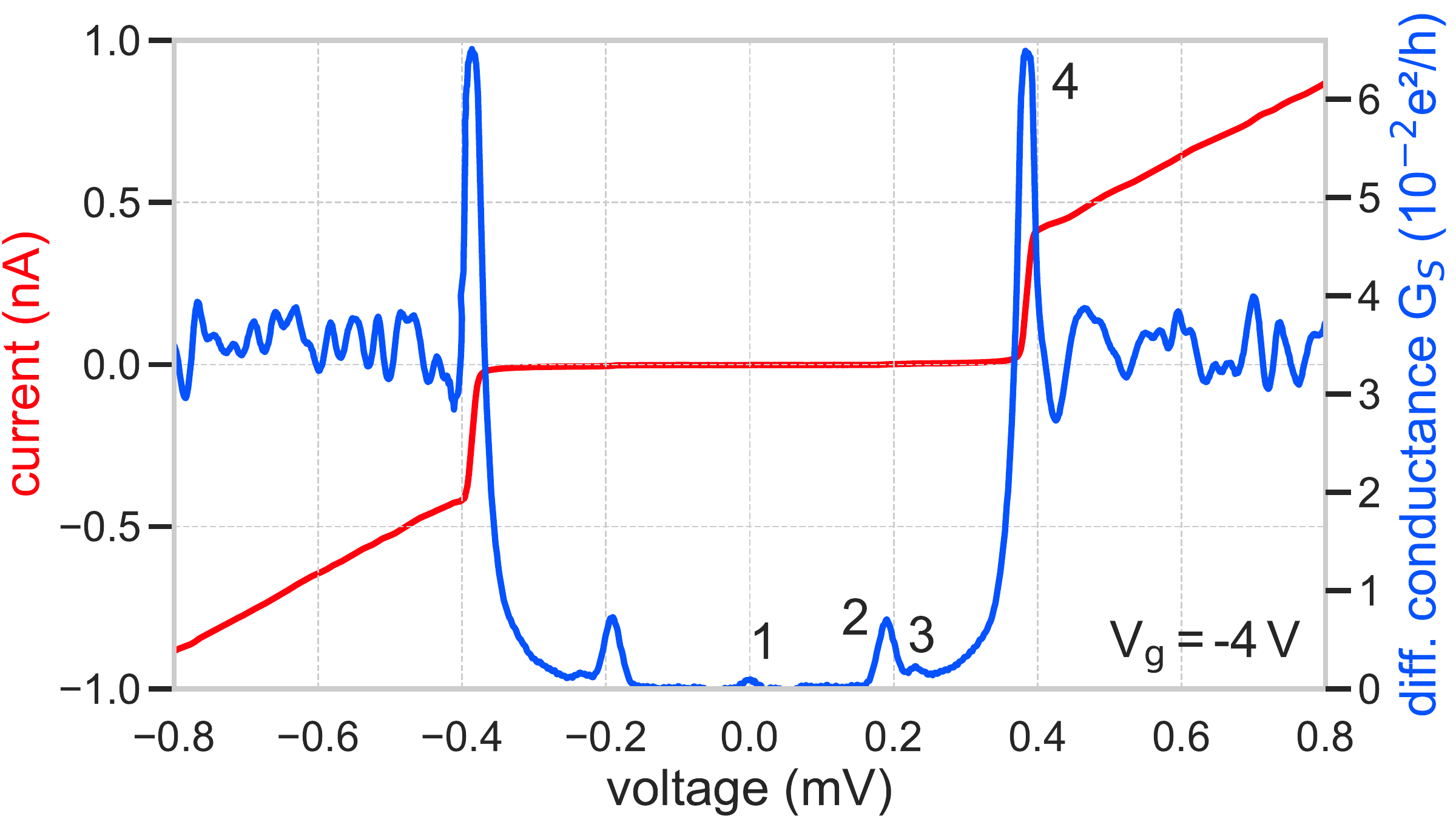}
\caption{Voltage-driven tunnel spectroscopy measurement of an InAs/Al nanowire junction at $V_g=-4$\,V: Current as a function of voltage (red) and corresponding differential conductance in units of $e^2/h$ (blue). The four peaks correspond to (1) the superconducting state, (2)$\,$(3) sub-gap structures and (4) the induced proximity gap. 
}
\label{fig:hard_gap}
\end{figure}
\subsection{Hard gap spectroscopy and Yu--Shiba--Rusinov states}

The second benchmark of a superconductor-semiconductor hybrid structure is the hardness of the induced proximity gap, i.e. the coupling strength between the nanowire and the metal shell, in the single-channel or tunneling limit. Previous experiments on similar systems used tunnel spectroscopy between one superconducting and one normal conducting contact to detect the strong change in the density of states of the device close to the gap edge \cite{Chang15}. Even though our structure is slightly different in terms of the contact setup, we can obtain comparable conditions by applying a negative gate voltage which sets the device into the tunnel-limited regime \cite{Guel17}. Figure~\ref{fig:hard_gap} shows a typical spectroscopy-like measurement of a self-defined InAs/Al nanowire Josephson junction. 
\begin{figure*}[t]
\centering
\includegraphics[width=0.9\textwidth]{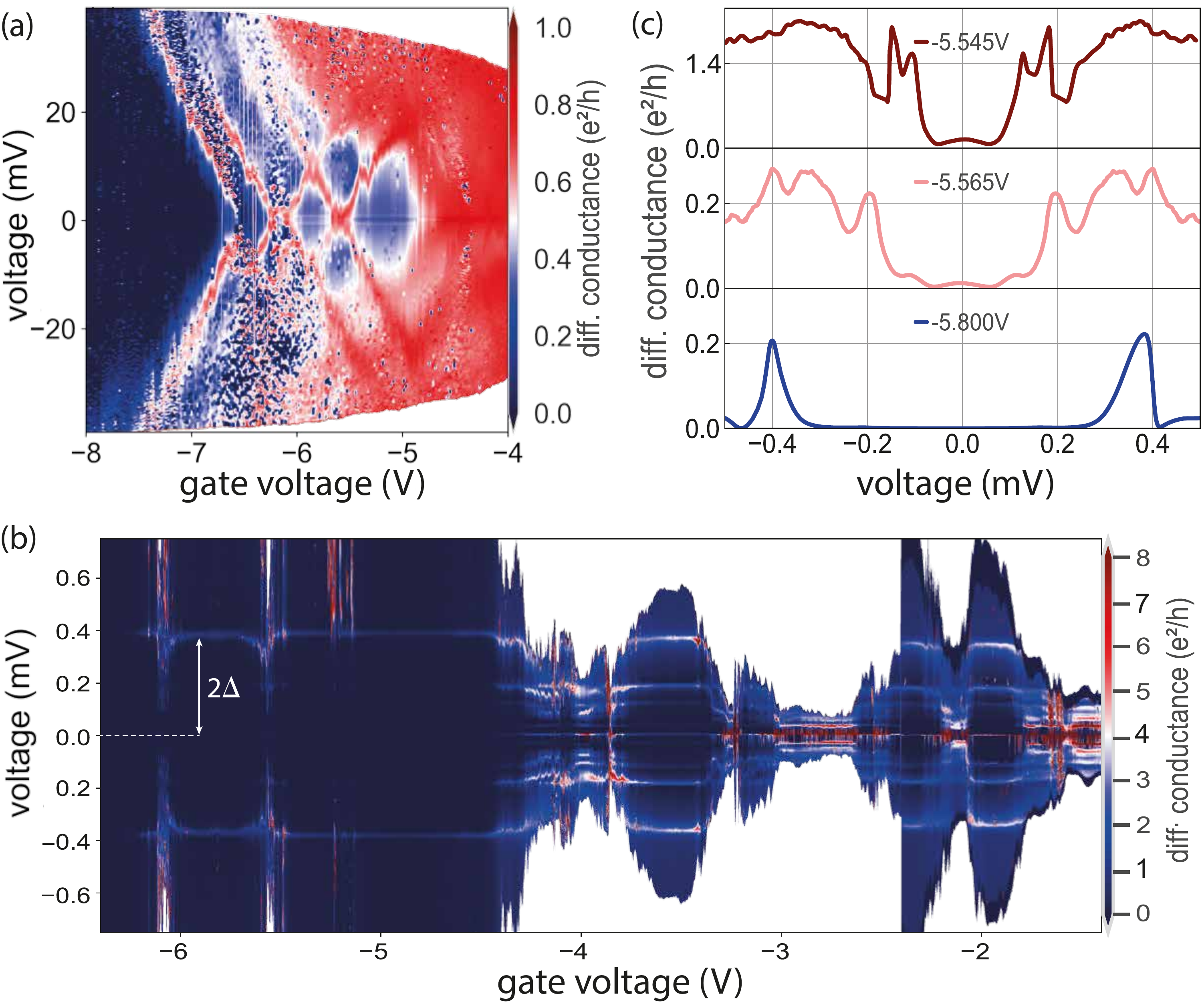}
\caption{(a) Full gate response for gate voltages between $-8$ and $-4\,$V, showing a set of Coulomb diamonds which are related to a single quantum dot, with a charging energy of $E_c=11.6\,$meV$\,\approx\,60\Delta$. (b) 
Differential conductance vs. bias voltage measurements with a high resolution gate voltage stepping for $V_g\,=-6.4 \dots-1.4$\,V. The measurement covers 
the in- and inter-diamond section as well as the open-junction-regime. Within the Coulomb diamonds found in the range of $V_g<5.5$\,V, the only visible feature is related to the edge of the proximity gap. However, for voltage values close to the section in which the quantum dot is open, Yu--Shiba--Rusinov states are observable. (c) Differential conductance vs. bias voltage traces at different gate voltages, depicting the evolution of the sub-gap structure in and out of the Coulomb diamonds. For $-5.8\,$V (dark blue), there are only two peaks which are related to the edge of the proximity gap. At  $-5.655\,$V (light red), multiple sub-gap states become visible. For $-5.545\,$V (dark red), the spectrum completely changes, ending up in a trace with a 7 times larger conductance and a pronounced peak at zero energy. We attribute this behavior to a mixture of both multiple Andreev reflections and Yu--Shiba--Rusinov states.}
\label{fig:gate_combined}
\end{figure*}
As indicated in the figure, four pronounced peaks are identified. Feature (1) is thereby related to the superconducting state, while (2) and (3) correspond to sub-gap structures. Finally, the dominating peak (4) marks the edge of the induced proximity gap 2$\Delta=380\,\mu$eV. However, more important than the existence of pronounced spikes in the conductance is the ratio of the conductance for in- and out-of-gap transport, i.e. $G/G_N$. Between (4) and (2), the conductance drops to $G/G_N\approx0.02$, with a further reduction between (2) and (1) to $G/G_N<0.001$. Such a strong decrease of the conductance is an indication that our devices host a so-called hard gap \cite{Guel17,Chang15,Carrad19}.

As already shown in Fig.~\ref{fig:gatesweep}, the actual junction behavior is strongly correlated with the applied gate voltage. Thus, we performed a gate batch measurement between $-8$ and $0\,$V in steps of $29\,$mV, which covers both the tunnel-limited regime as well as the range in which the transport is dominated by multiple Andreev reflections. Figure \ref{fig:gate_combined} (a) shows the stability diagram of the junction for gate voltages between $-8$ and $-4$\,V. Here, the transport is dominated by single electron tunneling, which results in pronounced Coulomb diamonds. By mapping out the size of the diamonds in terms of their individual gate and bias voltage, we obtain a lever arm $\alpha=0.095$ and charging energy $E_c=11.6\,$meV. Based on these values, especially due to $\Delta < E_c$ \cite{Krisanskas15}, we assume that the transport, in the sections in which the quantum dot is open, is mediated by Yu--Shiba--Rusinov (YSR) states rather than by conventional Andreev bound states \cite{Yu75,Shiba68,Rusinov69}. 
\begin{figure*}[t]
\centering
\includegraphics[width=0.98\textwidth]{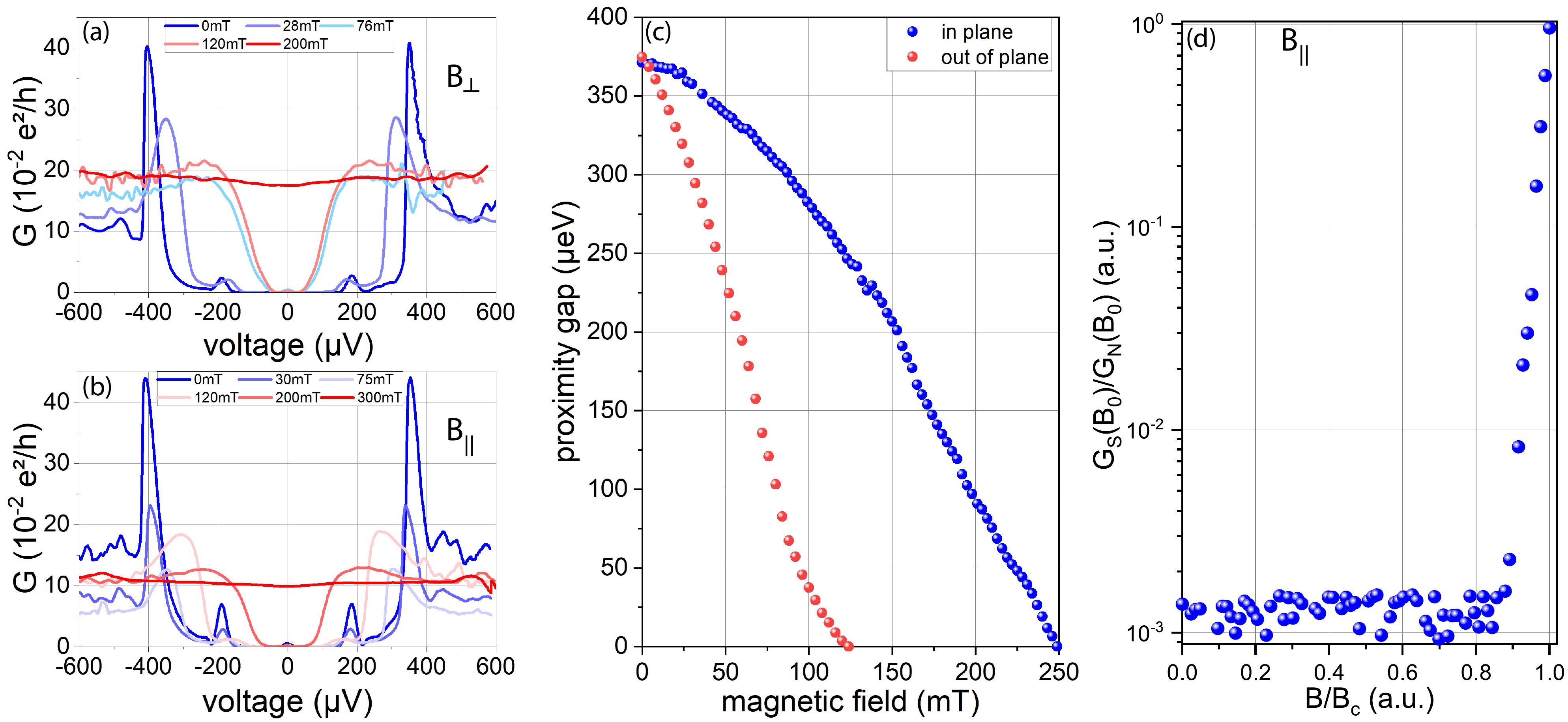}
\caption{Differential conductance traces as a function of bias voltage for various magnetic fields both perpendicular (a) and parallel (b) to the nanowire axis. For larger field values, the peaks corresponding to the edge of the proximity gap are damped and move closer to zero energy. (c) Width of the proximity gap 2$\Delta$ for an increasing magnetic field between 0 and 250$\,$mT. The energies were extracted by using the first derivative of the differential conductance. For an out-of-plane orientation, we find a critical field of $B_c\,$=130$\,$mT, while for the in-plane field we obtain $B_c\,$=250$\,$mT.  (d) Ratio of the averaged in-plane-field-dependent zero bias conductance relative to the conductance above the critical field. As long as the field is less than 90$\%$ of the critical field, the zero bias conductance is mainly determined by the noise floor of the system, without any parasitic sub-gap states, which is another proof for the hard-gap behavior of the junction.}
\label{fig:magnetic_field_combined}
\end{figure*}
In order to probe the actual state structure, we performed differential conductance vs. bias voltage measurements at a high resolution gate voltage stepping in the range of $-6.4$ to $-1.4\,$V,
which is shown in Fig.~\ref{fig:gate_combined} (b).\footnote{Compared to the measurements shown in Fig.~\ref{fig:gate_combined} (a) the pattern is shifted slightly because of gate drift.}  We can distinguish between three different regimes. For less negative gate voltages, i.e. values above $-4.3\,$V, (multiple) sub-gap states arise. However, for more negative gate voltages and, especially inside of the Coulomb diamonds, the sub-gap transport is completely suppressed. Nevertheless, the edges of the induced proximity gap are still present, represented by two straight light-blue lines at a constant voltage of $\pm$375$\,\mu$V. On the other hand, for certain gate voltage ranges, in which the quantum dot is open, several sub-gap features arise. In Fig.~\ref{fig:gate_combined} (c) three exemplary curves in the range around $-5.7$\,V are shown. For a gate voltage of $-5.8$\,V, the junction characteristics exhibit a clear hard-gap without any sub-gap features. If the voltage increases, i.e. up to $-5.565\,V$, a single state moves into the gap and additional peaks appear, which are probably related to coherent reflection events of the same state. For a gate voltage of $-5.545\,V$, the state reaches zero energy, which, together with the onset of single electron tunneling, results in a 4 times increase of the in- and out-of-gap conductance as well as the emergence of a pronounced zero bias peak, probably caused by the interplay of multiple Andreev reflections and a single Yu--Shiba--Rusinov state. Based on  the strong gate asymmetry, as well as due to the relatively large charging energy of $E_c\approx 60\Delta$, this behavior could be an indication for a quantum phase transition from a singlet into a doublet state, which effectively changes the nature of the junction from "0" to "$\pi$" \cite{Li17,Jergensen07}. However, due to the fact that the junction does not carry a real, non-dissipative current in this specific gate voltage range and is not embedded in a phase-sensitive device like a superconducting quantum interference device, there is no direct way to probe the resulting reversal of the supercurrent. Nevertheless, these states could still offer some advantages for devices based on mesoscopic Josephson junctions by adding an additional energy-tunability of the states besides the superconductor-semiconductor interface transparency.

\subsection{In-plane and out-of-plane magnetic field transport measurements}

The common requirement for all devices based on Andreev bound states is the manipulation of the phase across the junction in order to move along the dispersion relation and set the system to a fixed operational point. This is normally achieved by means of a superconducting loop connecting the junction and applying a small external out-of-plane magnetic field which generates a flux-induced phase shift. In the case of semiconductor nanowires, and more specifically for materials like InAs or InSb, the strong spin-orbit coupling as well as the large g-factor can be used as an additional way to manipulate the state structure both in energy and momentum space. Here, an additional in-plane field is required to induce a Zeeman splitting, with field values which can easily overcome the bulk critical field of Al by one or two orders of magnitude. Both things together make the magnetic field sustainability the second key benchmark of a nanowire Josephson junction.

In order to probe the magnetic field stability of the in-situ junctions we performed voltage-driven measurements at a gate voltage of $-4.3\,$V, i.e. close to, but outside of the regime which is dominated by the quantum dot. Figure \ref{fig:magnetic_field_combined} (a) and (b) show the conductance vs. bias voltage for an out-of-plane field as well as for a magnetic field which is applied parallel to the nanowire axis, respectively. 

As can be seen, both measurements show the typical damping and smearing out effect of the gap edge peaks, which is related to the continuous closing of the induced proximity gap if the magnetic field is increased. The field-induced changes in the normal-state conductance are most likely related to quantum fluctuations in the nanowire, i.e. the previously mentioned universal conductance fluctuations, induced by the superposition of multiple, flux-penetrated scattering loops. However, these should just affect the background conductance rather than the transport below the superconducting gap. By using the first derivative of the differential conductance, it is possible to extract the actual gap width for each trace. Even though this method underestimates the gap width for small magnetic fields, it makes it possible to find reliable values when the actual gap edge peaks are already completely suppressed. Figure~\ref{fig:magnetic_field_combined} (c) summarizes the field-dependent gap width in terms of energy for both the in-plane (blue) as well as the out-of-plane (red) field. For the latter one, we obtain a critical field of $B_c=130\,$mT, while the first one is almost two times larger, with $B_c=250\,$mT. These comparably small values of $B_c$ with respect to other works can be attributed to the much thicker Al shell of 25$\,$nm in our case \cite{Meservey71,Krogstrup15}. Additionally, both traces shown in Fig~\ref{fig:magnetic_field_combined} (c) deviate from the conventional BCS-behaviour, resulting in a much weaker field-dependency of $\Delta$. However, we would like to stress that the closing of the gap is not necessarily coupled to a softening of the gap, as can be seen in Figure \ref{fig:magnetic_field_combined} (d). Here, we analyzed the change of the ratio between the averaged field-dependent zero bias conductance $G_S(B_0)$ and the conductance above $B_c$, $G_N(B_0)$, as a function of fractions of the critical field. As long as the applied in-plane field is smaller than $0.9 B_c$, $G_S(B_0)$ is mainly limited by the noise floor of the system. However, above this value, the conductance increases continuously, until it reaches $G_N(B_0)$. We interpret this as another proof for the hardness of the induced proximity gap in our system, which is not disturbed by parasitic sub-gap states even for comparably large magnetic fields. 

\section{Conclusion}
We have demonstrated that mesoscopic Josephson junctions based on nanowires with shadow-mask-defined weak links show state-of-the-art properties in terms of gap-hardness, gate tunability of the switching current, interface transparency and magnetic field resilience. In fact, in contrast to conventional nanowire junctions with epitaxial Al full or half shells, they provide much more flexibility for the usable superconductors due to the avoidance of a wet or dry chemical etching step. 
Additionally, caused by the coupling between the actual junction length and the nanowire diameter, they ease the fabrication of short junctions, i.e. systems with just a single Andreev bound state. Thus, these novel junctions can potentially act as a building block for quantum devices based on excitations, like the Andreev qubit or topological systems with Majorana zero modes, which require a precise control of the internal state structure of the junction.

\section*{Acknowledgment}
Assistance with the MBE growth by Christoph Krause is gratefully acknowledged. Dr. Stefan Trellenkamp is gratefully acknowledged for electron beam lithography. Dr. Gianluigi Catelani is gratefully acknowledged for theory support regarding the magnetic field measurements. All samples have been prepared at the Helmholtz Nano Facility \cite{GmbH2017}. The work at RIKEN was partially supported by Grants-in-Aid for Scientific Research (A) (No. 19H00867) and Scientific Research on Innovative Areas “Science of hybrid quantum systems” (No. 15H05867). This work was partly funded by the Deutsche Forschungsgemeinschaft (DFG, German Research
Foundation) under Germany's Excellence Strategy – Cluster of Excellence Matter and Light for Quantum Computing (ML4Q) EXC 2004/1 – 390534769. We also  gratefully acknowledge support from DFG project SCHA 835/8-1 and UK EPSRC grant EP/P000916/1. The work at Forschungszentrum J\"ulich was supported by the project "Scalable solid state quantum computing" financed by the Initiative and Networking Fund of the Helmholtz Association.

\bibliography{NW-InAs-in-situ-junctions}
\end{document}